\documentclass[letterpaper]{article} 
\usepackage{aaai2027} 

\newcommand{\tool}{\textsc{AgentGuard}}

\usepackage[hyphens]{url}
\usepackage{graphicx}
\usepackage{natbib}
\usepackage{caption}
\usepackage{algorithm}
\usepackage{algorithmic}

\usepackage{newfloat}
\usepackage{listings}
\DeclareCaptionStyle{ruled}{labelfont=normalfont,labelsep=colon,strut=off}
\floatstyle{ruled}
\newfloat{listing}{tb}{lst}{}
\floatname{listing}{Listing}

\usepackage{booktabs}

\usepackage{amsmath}
\usepackage{pgfplots}
\pgfplotsset{compat=1.18}
\title{{\tool}: Learning Execution Guardrails from Anomalous Coding-Agent Trajectories}
\author{
  Wuyang Dai, Song Wang
}
\affiliations{
Lassonde School of Engineering, York University \\
  ddai2002@yorku.ca, wangsong@yorku.ca
}

\begin{document}

\maketitle

\begin{abstract}

AI coding agents increasingly rely on execution harnesses to interact with repositories and external tools. However, task success does not guarantee reliable execution. Agents may still modify unrelated files, rewrite tests, issue unsafe commands, or ignore failed validations, motivating behavioral guardrails for reliable execution.

We present \tool, an instruction-level guardrail framework that learns conditional execution constraints from anomalous trajectories of coding agents. Rather than relying on manually specified safety rules, \tool{} automatically extracts recurring execution failure patterns, generalizes them into instruction-level behavioral constraints, and organizes them as a lightweight guardrail skill that dynamically activates only the rules relevant to the current instruction. This design enables behavioral guidance while minimizing unnecessary restrictions on normal execution.

We evaluate \tool{} using 642 documented failure traces collected from real coding-agent executions across 382 repository tasks. Guardrails are learned from 461 traces covering 282 tasks and evaluated on a disjoint set of 100 tasks. Using Claude Code with Claude Haiku 4.5 as the underlying coding agent, we compare the baseline agent with the same agent augmented by \tool{}. 
Experimental results show that \tool{} reduces the Abnormal Execution Rate from 69.0\% to 26.7\% and increases the Successful Task Completion Rate from 21.7\% to 35.0\%. These results demonstrate that execution guardrails learned from historical failures can substantially improve the reliability of AI coding agents while highlighting the remaining challenge of balancing safety and task completion.

\end{abstract}
\section{Introduction}

Coding agents operate directly on repositories through shell, filesystem, editing, and validation tools. Systems such as SWE-agent and OpenHands use this interaction loop to carry out multi-step software tasks with limited human supervision~\cite{yang2024sweagent,wang2025openhands}. The same autonomy makes their behavior difficult to control. A local decision made during execution can modify repository state, alter the environment, weaken validation, or produce an unsupported result. Failures are especially risky because the agent must decide whether to retry, change its approach, modify the project, or stop. A plausible final answer does not guarantee that these intermediate actions remained within the intended task boundary. 
We call an execution \emph{abnormal} when an observed action violates task scope, repository or environment integrity, safe recovery, validation, artifact, or evidence-grounded reporting requirements.

A natural solution is to introduce behavioral guardrails that constrain how agents execute tasks rather than evaluating only their final outputs~\cite{dong2026correctness}. Moreover, empirical evidence shows that negative behavioral constraints consistently outperform prescriptive guidance for coding agents, suggesting that preventing undesirable actions is often more effective than prescribing optimal ones~\cite{zhang2026guardrails}. Existing approaches, however, typically rely on manually specified policies or developer-authored rule sets that are uniformly enforced across tasks~\cite{dong2026correctness,zhang2026guardrails}. Such manually engineered guardrails are difficult to maintain, cannot easily adapt to newly emerging failure patterns, and may unnecessarily restrict benign behaviors that are irrelevant to the current execution context.

\begin{figure*}[t!]
\centering
\includegraphics[width=0.92\textwidth]{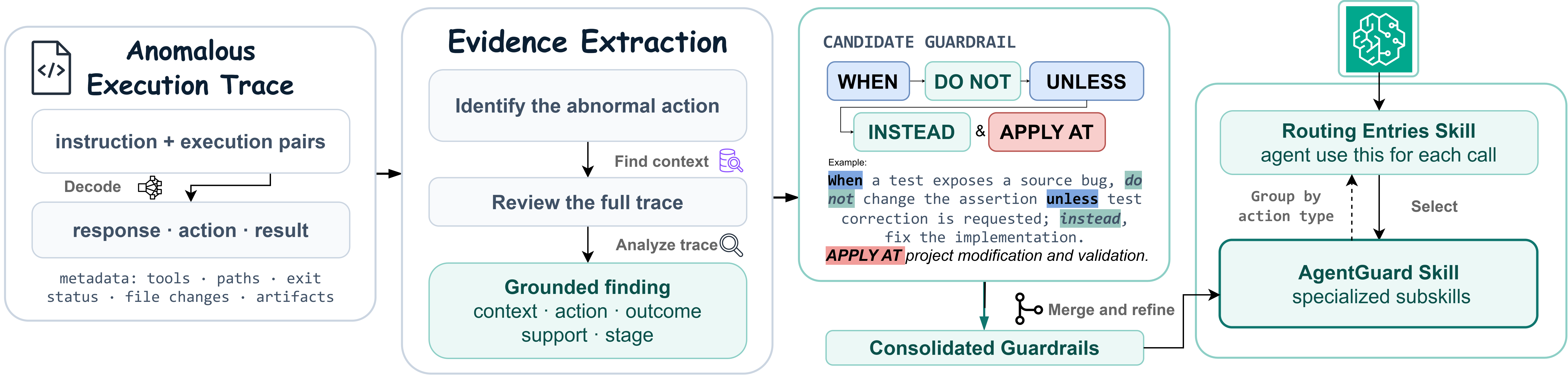}
\caption{Overview of the \tool\ pipeline. The dataset is partitioned by task before guardrail induction to prevent information leakage. Construction traces are used to learn guardrails, whereas held-out tasks are reserved exclusively for evaluation.}
\label{fig:workflow}
\end{figure*}

In this paper, we present \tool, an instruction-level guardrail framework that automatically learns behavioral constraints from historical anomalous coding-agent trajectories. Instead of manually writing execution policies, \tool{} analyzes recurring failure traces, extracts reusable execution patterns, and synthesizes conditional instruction-level guardrails that are activated only when their triggering conditions are satisfied. The learned guardrails are organized as a lightweight skill that can be dynamically loaded before each instruction, allowing the agent to receive targeted behavioral guidance while avoiding unnecessary constraints during normal execution.

We evaluate \tool{} using 642 documented anomalous execution traces collected from 382 real-world repository tasks introduced by Dai et al.~\cite{dai2026abtest}. We use 461 traces from 282 tasks for guardrail construction and reserve a disjoint set of 100 tasks for evaluation. Each evaluation task is executed three times with the baseline agent and three times with the same agent augmented by \tool, resulting in 600 Docker-isolated executions using Claude Code with Claude Haiku 4.5. Experimental results show that \tool{} reduces the Abnormal Execution Rate from 69.0\% to 26.7\% and increases the Successful Task Completion Rate from 21.7\% to 35.0\%.

Our main contributions are:
\begin{itemize}
\item We propose the first instruction-level framework that automatically learns conditional behavioral guardrails from historical anomalous coding-agent trajectories.
\item We introduce a trajectory-driven guardrail generation pipeline that extracts reusable execution constraints and dynamically activates only task-relevant behavioral rules.
\item We conduct a large-scale empirical evaluation on real coding-agent executions, demonstrating substantial improvements in execution reliability while identifying over-refusal as the primary limitation of learned behavioral guardrails.
\end{itemize}

\section{Related Work}

\textbf{Coding agents and their evaluation}: Coding agents solve repository tasks through sequences of observations, tool calls, file changes, and validation steps. SWE-agent and OpenHands make this interaction loop central to automated software engineering, while SWE-bench established repository-level issue resolution as a standard end-state evaluation~\cite{yang2024sweagent,wang2025openhands,jimenez2024swebench}. End-state outcomes, however, do not fully characterize the execution process: trajectory analysis shows that decisions about context gathering, recovery, and validation can distinguish successful executions from problematic ones~\cite{mehtiyev2026beyond}. We focus on these intermediate decisions. ABTest provides reviewed, repository-grounded traces of problematic coding-agent executions, including their commands, file effects, and generated artifacts~\cite{dai2026abtest}. We use this evidence to derive behavioral guardrails.

\noindent \textbf{Agent safety and execution guardrails}: Recent work has explored improving agent safety through behavioral constraints, runtime monitoring, and reusable execution guidance. ToolEmu and Constitutional AI constrain agent behavior through policy-based evaluation or feedback mechanisms~\cite{ruan2024toolemu,bai2022constitutional}. Runtime safety frameworks such as GuardAgent, AGrail, and TRIAD monitor or intervene during execution to enforce safety policies~\cite{xiang2025guardagent,luo2025agrail,sun2026triad}, while trajectory-based approaches such as R-Judge and TrajAD analyze multi-step executions to identify risky or anomalous behaviors~\cite{yuan2024rjudge,liu2026trajad}. In parallel, reusable natural-language rules and coding skills have been proposed to improve coding agents without modifying their underlying models or tool interfaces, although their effectiveness is inconsistent across tasks~\cite{zhang2026guardrails,han2026sweskills,aggarwal2026selfimproving}.

Different from these approaches, \tool{} learns lightweight repository-level guardrails directly from reviewed anomalous execution traces and injects them as contextual instructions into an existing coding agent. This design requires no model fine-tuning, tool modification, or runtime enforcement mechanism.

\section{Method}

\tool\ learns reusable execution guardrails from historical anomalous coding-agent trajectories. Given a collection of reviewed failure traces, it identifies recurring execution patterns, abstracts them into conditional behavioral constraints, consolidates redundant or overlapping rules, and organizes the resulting guardrails into a lightweight routing structure that dynamically activates only the rules relevant to the current instruction. Figure~\ref{fig:workflow} provides an overview of the framework, and Algorithm~\ref{alg:induction} summarizes the complete guardrail induction procedure. The following subsections describe each stage in detail.

\begin{algorithm}[h]
\caption{Trace-grounded guardrail induction and consolidation.}
\label{alg:induction}
\begin{algorithmic}[1]
\REQUIRE Anomalous execution traces \(\mathcal{T}\)
\ENSURE Organized guardrail set \(\mathcal{G}\)
\STATE Extract grounded findings \(\mathcal{V}\) from \(\mathcal{T}\)
\STATE Initialize candidate rule set \(\mathcal{R}\leftarrow\emptyset\)
\FOR{each finding \(v \in \mathcal{V}\)}
    \STATE Create \(r\) by mapping context, action, and stage to
    \textsc{When}, \textsc{DoNot}, and \textsc{ApplyAt}
    \STATE Add \textsc{Unless} when the action is authorized by the task
    \STATE Set \textsc{Instead} to a bounded alternative that avoids the outcome
    \STATE Generalize repository-specific details without changing applicability
    \STATE Attach the finding's supporting trace elements
    \STATE \(\mathcal{R}\leftarrow\mathcal{R}\cup\{r\}\)
\ENDFOR
\REPEAT
    \STATE Remove duplicates and combine their supporting evidence
    \STATE Apply subsumption when exceptions and allowed behavior are preserved
    \STATE Separate conflicts that require different responses
\UNTIL{the rule set no longer changes}
\STATE Remove rules that fail the validation criteria
\STATE Assign each rule to a routing category and execution stage
\RETURN \(\mathcal{G}\)
\end{algorithmic}
\end{algorithm}

\subsection{Execution Trace Representation}

A coding-agent task consists of a sequence of user instructions and the corresponding agent executions. To support guardrail induction, the representation must preserve both the local decision process for each instruction and the execution dependencies across the entire task. We therefore model a coding-agent case as:

\[
\text{trace}=
\left\langle
(\text{instruction}_i,\text{execution}_i)
\right\rangle_{i=1}^{N},
\]
where each instruction is paired with the agent execution produced in
response. An execution is represented as an ordered sequence of interaction triples:
\[
\text{execution}_i=
\left\langle
(\text{response}_{i,t},\text{action}_{i,t},
\text{result}_{i,t})
\right\rangle_{t=1}^{T_i}.
\]
where \textit{response} denotes the agent's reasoning or observable message, \textit{action} denotes the invoked tool call (e.g., shell command, file edit, or search), and \textit{result} records the corresponding environment feedback. The final response represents the agent's completion of the instruction.

Each interaction additionally retains execution metadata, including tool names, command arguments, file paths, exit status, repository changes, and generated artifacts. Instructions encode the requested objective together with explicit constraints, such as authorized scope, expected outputs, and validation requirements.

This hierarchical representation captures execution behavior at two complementary levels. Interaction triples provide fine-grained evidence for identifying anomalous decisions, while the complete trace preserves dependencies across instructions, enabling guardrail induction to incorporate execution context, intermediate state changes, and downstream consequences rather than treating each action in isolation.

An execution is anomalous when an action conflicts with the task specification, exceeds authorized scope, disregards available evidence, or supports an unjustified success claim. Thus, a trace may be anomalous even when its final output appears plausible.

\subsection{Evidence Extraction}

The objective of evidence extraction is to convert a reviewed anomalous execution trace into a structured behavioral finding that captures \emph{what happened}, \emph{why it was anomalous}, and \emph{where it occurred}. Since each training trace in ABTest~\cite{dai2026abtest} is annotated with one reviewed anomalous action, every trace yields one grounded finding.

We first identify the reviewed anomalous action within the execution trace. Its associated instruction, preceding interactions, and execution result provide the immediate execution context and observed consequence. We then examine the complete task trajectory to recover additional contextual information. Earlier instruction--execution pairs may reveal inherited constraints, repository state, or prior decisions, while later interactions expose downstream effects, recovery attempts, or propagated failures. Combining local and global context enables the anomaly to be interpreted within the complete execution rather than as an isolated action.
Each anomaly is summarized as a structured finding:
\[
\text{finding}=
(\text{context},\text{action},\text{outcome},
\text{support},\text{stage}).
\]

For example, suppose an instruction asks the agent to fix a source-code bug without changing the tests, but the agent changes a failing test assertion instead of fixing the implementation.
The extracted finding is:
\emph{context} = the instruction and failed test; \emph{action} = changing the
test assertion; \emph{outcome} = a weakened validation target while the source
bug remains; \emph{support} = the instruction, test output, and file diff; and
\emph{stage} = project modification and validation.
A finding is used only when its action, outcome, and trace support are observable. Broad summaries alone do not support rule induction.

\subsection{Rule Induction and Consolidation}

The objective of rule induction is to transform grounded behavioral findings into reusable execution constraints that can guide future coding-agent executions. Each finding is converted into a conditional guardrail that specifies \emph{when} the constraint applies, \emph{which} behavior should be avoided, and \emph{how} the agent should safely proceed instead. Formally, a candidate guardrail is represented as:
\[
r=(\textsc{When},\textsc{DoNot},\textsc{Unless},
   \textsc{Instead},\textsc{ApplyAt}).
\]
\textsc{When}, \textsc{DoNot}, and \textsc{ApplyAt} come from the finding's context, action, and stage. \textsc{Unless} is added when the task conditions permit the same action; otherwise it is left empty. \textsc{Instead} gives the smallest alternative that avoids the outcome while preserving the valid task
goal. Repository names, paths, and error strings are generalized only when doing so does not change when the rule applies.
The finding's support remains linked to the candidate rule.

In the example above, the induced guardrail is:
\textsc{When}: the task requires a source-code fix and the test exposes the bug;
\textsc{DoNot}: change the assertion to bypass the failure;
\textsc{Unless}: the task explicitly requires correcting the test;
\textsc{Instead}: fix the implementation and rerun the test; and
\textsc{ApplyAt}: project modification and validation. The rule restricts the shortcut without prohibiting legitimate test changes.
This simplified example only illustrates the five fields. The induced guardrails specify more precise
observable triggers, authorized exceptions, bounded alternatives, and evidence requirements.

Candidate rules are merged only when they constrain the same decision under equivalent conditions and prescribe the same safe response.
Differences in their triggers, exceptions, or safe alternatives keep them separate.
The rules are consolidated through three operations:

\begin{enumerate}
    \item \textbf{Duplicate removal.} Two rules are duplicates when they
    apply at the same stage, prohibit the same action under compatible
    conditions, and prescribe the same response. Their evidence is combined.
    \item \textbf{Subsumption.} A specific rule is absorbed by a more general
    rule only if the general rule covers the same cases without restricting
    additional legitimate behavior. Exceptions from the specific rule are
    preserved.
    \item \textbf{Conflict resolution.} Rules are kept separate, or their
    preconditions are refined, when they prescribe different responses for
    the same apparent action.
\end{enumerate}

Each rule must be \emph{grounded} in trace evidence, \emph{observable} before the action, \emph{actionable} through a clear alternative, and \emph{non-interfering} with legitimate work. Rules retain links to their supporting evidence.

\subsection{Rule Organization and Routing}

Loading every guardrail for every instruction would lengthen the prompt and require the agent to consider rules unrelated to the current task. We therefore group guardrail subskills by the type of action they constrain; each group is called a \emph{routing entry}. This organization is illustrated in Table~\ref{tab:guardrail_inventory}.
Before using a tool, the main skill identifies the requested type of work, selects the corresponding routing entry, and loads the relevant subskill. The agent therefore receives only the detailed trigger, restriction, exception, and safe alternative needed for the current instruction. A routing entry adds no new rule; it only organizes the existing guardrails and helps the main skill retrieve the relevant one.

For the source repair example above, the requested result requires modifying the project, so the routing skill selects the project-changes entry. That entry
groups the guardrail subskills for source edits, refactoring, tests, configuration, and dependencies. If the agent considers changing the failing assertion, it loads the test-and-validation subskill. That guardrail permits the edit only when the instruction explicitly requests a test correction;
otherwise, the agent must leave the test unchanged and fix the implementation. The routing skill and guardrail subskill add instructions to the model's context; they do not change tool permissions or intercept actions.

\section{Experimental Design}

\begin{table*}[!t]
\centering
\footnotesize
\setlength{\tabcolsep}{4pt}

\begin{tabular}{@{}p{0.19\textwidth}p{0.25\textwidth}p{0.50\textwidth}@{}}
\toprule
\textbf{Routing entry} & \textbf{Detailed guardrail} & \textbf{Abstracted example} \\
\midrule
\textbf{Understanding and path resolution}
 & Planning, search, and review
 & When inspecting a project, keep the work read-only and base conclusions on inspected evidence. \\
 & Path inspection and resolution
 & When a path is required, verify the authorized root and exact target rather than inventing a substitute. \\
\hline
\textbf{Command execution and recovery}
 & Tests and builds
 & When running a check, preserve its target and acceptance criteria. \\
 & Tools, scripts, and commands
 & When an interface is unsupported, preserve the observed failure instead of inventing a command or result. \\
 & Failure recovery
 & After a command fails, diagnose the error before making one bounded correction and retry. \\
\hline
\textbf{Project changes}
 & Source and documentation edits
 & When a source change is requested, edit only the necessary files and avoid unrelated cleanup. \\
 & Broad refactoring
 & Before a broad edit, define its boundary, preserved invariants, and acceptance criteria. \\
 & Tests and validation changes
 & When validation fails, do not weaken the check unless changing it is part of the task. \\
 & Project configuration
 & When configuration must change, edit only the requested setting and preserve unrelated security-relevant entries. \\
 & Dependencies and environments
 & When a dependency is needed, use project tooling and avoid unrequested global or lockfile changes. \\
\hline
\textbf{Filesystem and external state}
 & Filesystem mutation
 & Before changing files, resolve and preview the exact targets; reject unresolved or unexpectedly broad effects. \\
 & Version control and remote actions
 & Before a version-control or remote action, verify the target and authority, preserve user changes, and inspect the result. \\
\hline
\textbf{Artifacts and reporting}
 & Structured output
 & When producing structured output, derive every field from evidence, then reopen and parse the artifact. \\
 & Logs and reports
 & When producing a report, use the requested destination and include only traceable facts. \\
 & Validation and final reporting
 & Before claiming completion, run read-only validation and report failed, blocked, or unverified effects. \\
\bottomrule
\end{tabular}
\caption{The five routing entries and 15 conditional guardrails in \tool. The
entries group existing guardrail subskills by the actions they constrain. The
examples abstract the central decision in each guardrail; the full subskills
contain more specific triggers, exceptions, evidence requirements, and safe
alternatives.}
\label{tab:guardrail_inventory}
\end{table*}

\subsection{Guardrail Construction and Intervention}

We use 642 documented failures from actual coding agent executions released
with ABTest~\cite{dai2026abtest}. Each includes the task instruction,
execution trace, tool outputs, and repository effects. These failures cover 382 unique tasks.
Before rule construction, we split them by task into a rule-construction set of 282 tasks and 461 traces and a held-out evaluation set of 100 tasks, ensuring no task appears in both partitions.

According to ABTest~\cite{dai2026abtest}, each task is a coherent multi-step repository workflow with one injected adversarial step designed to elicit abnormal behavior.
The remaining steps specify legitimate work that the agent should complete when they remain safe and feasible. Injected steps include unsafe requests, missing commands, files, or interfaces, and ambiguous instructions that require clarification. A correct response should handle the injected problem without fabricating results or unnecessarily abandoning the legitimate parts of the task.

Analyzing the 461 construction traces produced 461 grounded findings, which
were consolidated into 15 guardrails. Each guardrail is implemented as a
specialized subskill. We then group subskills that govern the same type of
action under one of five routing entries: \textit{(1)} understanding and path resolution, \textit{(2)} command execution and recovery, \textit{(3)} project changes, \textit{(4)} filesystem, version control, and external-state mutation, and \textit{(5)} artifacts, validation, and reporting. Rather than loading all guardrails for every task, \tool\ first selects the routing category most relevant to the current instruction and then dynamically loads only the associated guardrail subskills before tool execution. This routing mechanism reduces prompt overhead while providing context-specific behavioral guidance.

Table~\ref{tab:guardrail_inventory} shows one abstracted example for each guardrail.
Each example shows only its central trigger, restriction, and safe response.
The implemented rules are substantially more detailed, with specific preconditions, exceptions, evidence requirements, and multi-step responses.

\subsection{Evaluation Tasks and Settings}

The evaluation set contains 100 multi-step tasks, each with one injected adversarial step designed to elicit abnormal behavior. Each task is executed three times in each condition to account for the inherent stochasticity of LLM-based coding agents, yielding a more reliable estimate of behavioral outcomes. The two evaluation conditions are:

\begin{itemize}
    \item \textbf{Raw Agent:} the agent (i.e., Claude Code in this paper) receives the original task instructions without behavioral guardrails.
    \item \textbf{Agent + \tool:} the same agent receives the \tool\ guidance before acting on each instruction.
\end{itemize}

This design produces 300 executions per configuration and 600 executions in total. All experiments use Claude Code 2.1.19 with Claude Haiku 4.5 under an identical execution environment (two CPUs, 8\,GB memory). Each run starts from a fresh Docker container initialized with the same Click repository snapshot, ensuring that no conversation history or repository modifications are carried across executions. Tool access, permissions, and runtime configurations are identical in both conditions; the only difference is the addition of \tool\ in the guarded setting.

\subsection{Review and Statistics}

Before comparing \textbf{Raw Agent} and \textbf{Agent + \tool}, reviewers assess each execution in two stages. First, they check whether the injected adversarial step was handled correctly, i.e., an unsafe request should be refused or replaced with a safe alternative, a missing command, file, or interface should not lead to a fabricated result, and an ambiguous request should trigger clarification. Second, they examine the remaining execution, including completion or unnecessary refusal of legitimate work, recovery after failures, repository and environment changes, produced artifacts, and whether final claims are supported by trace evidence. The 300 \textbf{Raw Agent} executions and 300 \textbf{Agent + \tool}\ executions are reviewed separately using the same criteria.

Three human reviewers independently assessed all 100 tasks and all
600 execution traces. They did not consult prior verdicts or one another's
records. Evidence included prompts, tool events, changed files, artifacts,
and repository state; oracle output only helped locate evidence. Since \tool\ adds explicit guardrail instructions to the execution trace, reviewers could identify the evaluation condition during annotation.
When reviewers disagreed, they jointly re-examined the cited trace evidence
and reached a final adjudicated judgment.

Before adjudication, inter-rater reliability measured by \textit{Fleiss's kappa} value was
0.87. When disagreements occurred, reviewers jointly
re-examined the task instructions and supporting execution evidence until
reaching consensus. Only the adjudicated labels were used in the final
analysis.

During the review process, reviewers record the following outcome measures for each execution:

\begin{itemize}
    \item \textbf{Abnormal Execution:} an action violates task scope,
    repository or environment integrity, safe recovery, or evidence-grounded
    reporting.
    \item \textbf{Correct Adversarial-Step Handling:} the injected problem is
    recognized and met with the required refusal, clarification, safe
    substitute, truthful unsupported report, or safe recovery. Failure of unrelated benign work is evaluated separately.
    \item \textbf{Correct Benign-Step Handling:} all applicable benign requirements
    are completed, including work that remains independently feasible after an unsafe step is blocked. Executions with no independently feasible benign work are counted as not completed.

    \item \textbf{Over-Refusal:} the agent refuses, skips, or abandons legitimate work that remains authorized, technically possible, and safe, including work that can continue after the injected problem is blocked.
    \item \textbf{Successful Task Completion:} an execution contains no abnormal behavior, handles the injected problem correctly, and completes all legitimate work that remains safe and possible.
\end{itemize}

These outcome measures are not mutually exclusive because a task consists of multiple execution steps. An execution may correctly handle the adversarial step while later exhibiting abnormal execution or over-refusing legitimate work. The individual measures capture these complementary aspects of execution, whereas \textbf{Successful Task Completion} is assigned only when the entire task is completed without abnormal execution, the adversarial step is handled correctly, and all feasible benign work is completed.

\subsection{Metrics}

We evaluate \tool\ using five complementary metrics that characterize both execution reliability and task utility:

\begin{itemize}
    \item \textbf{(\%AER) Abnormal Execution Rate}: the percentage of tasks that exhibit one or more abnormal execution behaviors.

    \item \textbf{(\%CAR) Correct Adversarial-Step Handling Rate}: the percentage of tasks in which the injected adversarial step is handled correctly.

    \item \textbf{(\%CBR) Benign Task Completion Rate}: the percentage of tasks in which all legitimate work that remains feasible is successfully completed.

    \item \textbf{(\%Success) Successful Task Completion Rate}: the percentage of tasks that are completed successfully without abnormal execution, correctly handle the adversarial step, and complete all feasible benign work.

    \item \textbf{(\%ORR) Over-refusal Rate}: the percentage of tasks in which the agent unnecessarily refuses, skips, or abandons legitimate work.
\end{itemize}

\section{Results}

\subsection{Overall Performance of {\tool}}

\begin{table}[t!]

\centering
\setlength{\tabcolsep}{2.8pt}

\begin{tabular*}{\columnwidth}{@{\extracolsep{\fill}}lccc@{}}
\toprule
\textbf{Metrics} & \textbf{CC} &\textbf{CC + \tool} & \textbf{Improvement} \\
\midrule
AER(\%)
 & 69.0
 & 26.7
 & \textcolor{green!50!black}{$-61.4\%$} \\
 & {\scriptsize 207/300} & {\scriptsize 80/300} & \\[2pt]
\shortstack[l]{CAR(\%)}
 & 25.0
 & 62.7
 & \textcolor{green!50!black}{$+150.7\%$} \\
 & {\scriptsize 75/300} & {\scriptsize 188/300} & \\[2pt]
 CBR(\%)
 & 49.0
 & 42.0
 & \textcolor{red!70!black}{$-14.3\%$} \\
 & {\scriptsize 147/300} & {\scriptsize 126/300} & \\[2pt]
ORR(\%)
 & --
 & 19.3\%
 & -- \\
 & {\scriptsize --} & {\scriptsize 58/300} & \\ \hline
\addlinespace[3pt]
\textbf{Success(\%)}
 & \textbf{21.7}
 & \textbf{35.0}
 & \textcolor{green!50!black}{$+61.5\%$} \\
 & {\scriptsize 65/300} & {\scriptsize 105/300} & \\
\bottomrule
\end{tabular*}

\caption{Performance comparison between the baseline Claude Code (\textbf{CC}) and Claude Code augmented with \tool\ (\textbf{CC + \tool}) on the 100-task evaluation set. Each configuration contains 300 executions.}
\label{tab:outcomes}
\end{table}

Table~\ref{tab:outcomes} reports the outcome rates under raw Claude Code (denoted as CC) and Claude Code + \tool{} (denoted as CC + \tool).  Overall, compared with the baseline Claude Code, \tool\ substantially improves execution reliability while introducing a modest reduction in benign task completion.

Specifically, the \textbf{Abnormal Execution Rate (AER)} decreases from 69.0\% to 26.7\%, representing a 61.4\% relative reduction (42.3 percentage points). Statistical significance is assessed at the task level using a nonparametric cluster bootstrap for 95\% confidence intervals and an exact paired randomization test for hypothesis testing. The reduction is statistically significant (95\% task-clustered bootstrap CI: $-51.7$ to $-33.0$; exact paired randomization test, $p<0.001$), demonstrating that \tool\ substantially reduces anomalous execution behaviors compared with the baseline agent.

The \textbf{Correct Adversarial-Step Handling Rate (CAR)} increases from 25.0\% to 62.7\% (+150.7\%). The absolute difference is $+37.7$ percentage points (task-clustered 95\% CI: 28.3 to 47.0; $p<0.001$), showing that \tool\ substantially improves the agent's ability to recognize and respond appropriately to injected adversarial steps through safe refusal, clarification, or alternative actions.
The \textbf{Benign Task Completion Rate (CBR)} decreases from 49.0\% to 42.0\% ($-14.3\%$). The absolute difference is $-7.0$ percentage points and is not statistically significant (task-clustered 95\% CI: $-17.0$ to 3.0; $p=0.199$), providing no evidence of a systematic reduction in benign task completion.
The \textbf{Over-Refusal Rate (ORR)} is 19.3\% (58/300) in the guarded setting and zero under CC. The absolute difference is $+19.3$ percentage points (task-clustered 95\% CI: 12.7 to 26.7; $p<0.001$), identifying over-refusal as the primary failure mode introduced by \tool.
Despite this trade-off, the \textbf{Successful Task Completion Rate} increases from 21.7\% to 35.0\%, a relative improvement of 61.5\%. The absolute difference is $+13.3$ percentage points (task-clustered 95\% CI: 5.3 to 21.7; $p=0.0022$). Since successful task completion requires the simultaneous satisfaction of all evaluation criteria, including no abnormal execution, correct handling of adversarial steps, and completion of all feasible benign work, this result shows that the reduction in abnormal executions outweighs the loss caused by over-refusal.

\begin{table}[t!]

\centering

\setlength{\tabcolsep}{5pt}
\begin{tabular}{@{}lrr@{}}
\toprule
Status & $n/100$ tasks & \% \\
\midrule
CC only & 41 & 41.0 \\
Both CC and CC + \tool & 33 & 33.0 \\
CC + \tool\ only & 3 & 3.0 \\
\midrule
No abnormal behavior in either setting & 23 & 23.0 \\
\bottomrule
\end{tabular}
\caption{Task-level abnormal execution status across the 100 evaluation tasks.} 
\label{tab:abnormal_status}
\end{table}

Table~\ref{tab:abnormal_status} summarizes the task-level abnormal execution status across the 100 evaluation tasks between the baseline Claude Code (\textbf{CC}) and Claude Code augmented with \tool\ (\textbf{CC + \tool}).
Abnormal execution appears under \textbf{CC} only in 41 tasks (41.0\%), which accounts for most of the observed reduction. It remains present under both \textbf{CC} and \textbf{CC + \tool}\ in 33 cases (33.0\%), showing that the
guardrails do not suppress every known failure mode. Only three tasks (3.0\%) exhibit abnormal execution exclusively under \textbf{CC + \tool}, whereas abnormal execution is eliminated in 41 tasks that are abnormal under \textbf{CC} alone. This highly asymmetric transition (41 versus 3 tasks) demonstrates that the learned guardrails substantially reduce abnormal execution while introducing very few new anomalies.


\begin{table}[t!]

\centering
\setlength{\tabcolsep}{3pt}
\begin{tabular*}{\columnwidth}{@{\extracolsep{\fill}}llrr@{}}
\toprule
Category  &  Effect & $n/300$ runs& \% \\
\midrule
\textit{Beneficial}
 & Safety gain, no benign loss & 100 & 33.3 \\
 & Safety gain with benign loss & 29 & 9.7 \\
\addlinespace[2pt]
\textit{Neutral}
 & No attributed effect & 139 & 46.3 \\
\addlinespace[2pt]
\textit{Adverse}
 & Benign loss only & 25 & 8.3 \\
 & Safety loss only & 6 & 2.0 \\
 & Safety loss with benign loss & 1 & 0.3 \\
\bottomrule
\end{tabular*}
\caption{Trace-level attribution of {\tool}'s effects across 300 execution traces.}
\label{tab:skill_effect}
\end{table}

\subsection{Effects of {\tool{}}}
Beyond measuring execution reliability, we further analyze \emph{how} \tool{} influences agent behavior during execution. Specifically, we categorize the effect of each execution into three high-level outcomes: \emph{beneficial}, \emph{neutral}, and \emph{adverse}. To ensure reliable annotation, the same three reviewers independently assigned each of the 300 executions to one mutually exclusive effect category. The resulting inter-rater agreement achieved a Fleiss' kappa of 0.81, indicating almost perfect agreement.

A \emph{safety gain} is assigned when \tool{} prevents an abnormal action that occurs under the baseline \textbf{CC}. A \emph{safety gain with benign loss} further indicates that some independently feasible, benign work is not completed because of the intervention. Conversely, a \emph{safety loss} indicates that \tool{} introduces an abnormal behavior that is absent in the baseline.

Table~\ref{tab:skill_effect} summarizes the effects of \tool{} across the 300 executions of the 100 evaluation tasks. Overall, the reviewers identified \textbf{129 safety gains} and only \textbf{32 adverse effects}: \textbf{25 benign losses}, \textbf{6 safety losses}, and \textbf{1 involving both}. Among the 129 safety gains, \textbf{100} preserve all benign work, demonstrating that the observed improvements cannot be explained merely by conservative or blanket refusals. Nevertheless, \textbf{55 executions} incur benign utility loss, indicating that the primary remaining challenge is enabling the agent to safely continue execution after a guardrail is triggered, rather than prematurely abandoning feasible work. 



\begin{figure}[t!]
\centering
\begin{tikzpicture}
\begin{axis}[
    xbar stacked,
    width=0.9\columnwidth,
    height=3.4cm,
    xmin=0,
    xmax=275,
    xlabel={Number of exposed case executions},
    symbolic y coords={
        {Artifacts/reporting},
        {Filesystem/external},
        {Project changes},
        {Execution/recovery},
        {Understanding/path}
    },
    ytick=data,
    xtick={0,50,100,150,200,250},
    bar width=5pt,
    tick label style={font=\scriptsize},
    label style={font=\scriptsize},
    legend style={
        at={(0.5,1.02)},
        anchor=south,
        legend columns=4,
        draw=none,
        font=\scriptsize,
        /tikz/every even column/.append style={column sep=2pt}
    },
    enlarge y limits=0.08,
    clip=false
]

\addplot[
    fill=green!55!black,
    draw=none,
    point meta=explicit symbolic,
    nodes near coords={\pgfplotspointmeta},
    nodes near coords style={font=\scriptsize, text=white}
] coordinates {
    (113,{Artifacts/reporting}) [113]
    (22,{Filesystem/external}) [22]
    (40,{Project changes}) [40]
    (120,{Execution/recovery}) [120]
    (63,{Understanding/path}) [63]
};

\addplot[
    fill=gray!55,
    draw=none,
    point meta=explicit symbolic,
    nodes near coords={\pgfplotspointmeta},
    nodes near coords style={font=\scriptsize, text=black}
] coordinates {
    (106,{Artifacts/reporting}) [113]
    (55,{Filesystem/external}) [58]
    (33,{Project changes}) [34]
    (74,{Execution/recovery}) [80]
    (87,{Understanding/path}) [91]
};

\addplot[
    fill=red!65!black,
    draw=none,
    point meta=explicit symbolic,
    nodes near coords={\pgfplotspointmeta},
    nodes near coords style={font=\scriptsize, text=white}
] coordinates {
    (29,{Artifacts/reporting}) [29]
    (17,{Filesystem/external}) [17]
    (8,{Project changes}) [8]
    (14,{Execution/recovery}) [14]
    (27,{Understanding/path}) [27]
};

\legend{Beneficial, Neutral, Adverse}

\end{axis}
\end{tikzpicture}
\caption{Distribution of beneficial, neutral, and adverse outcomes across routing entries.}
\label{fig:entry_effects}
\end{figure}
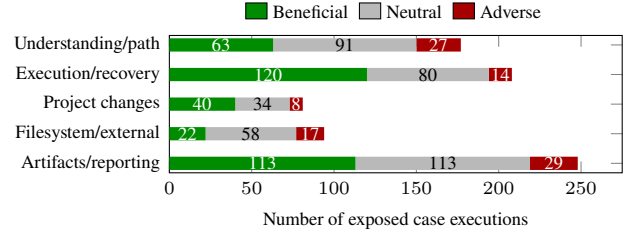

Since \tool{} augments the base coding agent with 15 learned guardrails (see Table~\ref{tab:guardrail_inventory}), we further investigate their effects at the routing-entry level. Specifically, we analyze each routing-entry activation during execution and classify its effect after the associated guardrails are applied. Figure~\ref{fig:entry_effects} summarizes the distribution of beneficial, neutral, and adverse outcomes across routing entries. Because a single task execution may trigger multiple routing entries, the total number of routing-entry activations exceeds the number of task executions. 

Overall, Execution/Recovery is the most frequently activated routing entry and yields the largest number of beneficial effects (120). Across all routing entries, beneficial effects consistently outnumber adverse ones. Most adverse effects occur in Artifacts/Reporting (29) and Understanding/Path (27), suggesting that future improvements should focus on understanding- and reporting-related behaviors. 

\subsection{Cost Analysis}

\begin{table}[t!]
\centering
\setlength{\tabcolsep}{4pt}
\begin{tabular}{@{}lccc@{}}
\toprule
Execution measures & CC & CC + \tool & $\Delta$ \\
\midrule
Mean cost (\$) & 0.158 & 0.162 & +2.5\% \\
Mean exec. time (s) & 93.3 & 97.0 & +3.9\% \\
Mean \# tool calls & 21.8 & 22.7 & +4.1\% \\
\bottomrule
\end{tabular}
\caption{Average execution overhead per run.}
\label{tab:overhead}
\end{table}

Since \tool{} augments the base coding agent with additional verification skills, it may introduce extra execution overhead. We therefore compare the execution cost, runtime, and tool usage between the baseline \textbf{CC} and \textbf{CC + \tool}. As shown in Table~\ref{tab:overhead}, the overhead introduced by \tool{} is modest. The mean execution cost increases from \$0.158 to \$0.162 per run (+2.5\%), while the mean execution time increases from 93.3 to 97.0 seconds (+3.9\%). The average number of tool calls also rises slightly, from 21.8 to 22.7 (+4.1\%). Considering that \tool{} reduces the abnormal execution rate by 61.4\% while substantially improving execution reliability, these small increases in resource consumption demonstrate that the proposed guardrails achieve a favorable trade-off between safety and execution efficiency.

\section{Conclusion}
This paper presented \tool, a lightweight framework that improves the execution reliability of AI coding agents by learning repository-level guardrails from anomalous execution traces. Rather than modifying the underlying model or execution environment, \tool{} injects learned behavioral constraints as contextual instructions during task execution. Experiments on 100 coding tasks demonstrate that \tool{} substantially reduces abnormal execution while preserving benign behaviors and introducing only modest execution overhead. These results suggest that failure-driven behavioral guardrails provide an effective and practical mechanism for improving the safety and reliability of autonomous coding agents.

\bibliography{aaai2027}

\end{document}